\documentclass[pra,aps,showpacs,twocolumn,floatfix]{revtex4}
\usepackage{epsf}
\usepackage[dvips]{color}
\newcommand {\be}{\begin{equation}}
\newcommand {\ee}{\end{equation}}
\newcommand {\bea}{\begin{eqnarray}}
\newcommand {\eea}{\end{eqnarray}}

\begin{document}

\title{Vortices and Ring Solitons in Bose-Einstein Condensates}
\author{L.~D. Carr$^{1,2}$ and Charles W. Clark$^{2}$\\}
\affiliation{1. Physics Department, Colorado School of Mines, Golden, CO 80401\\
2. Electron and Optical Physics Division, National Institute of
Standards and Technology, Technology Administration, U.S.
Department of Commerce, Gaithersburg, Maryland 20899}
\date{\today}

\begin{abstract}
The form and stability properties of axisymmetric and spherically
symmetric stationary states in two and three dimensions,
respectively, are elucidated for Bose-Einstein condensates. These
states include the ground state, central vortices, and radial
excitations of both. The latter are called ring solitons in two
dimensions and spherical shells in three.  The nonlinear Schrodinger
equation is taken as the fundamental model; both extended and
harmonically trapped condensates are considered.  It is found that
instability times of ring solitons can be long compared to
experimental time scales, making them effectively stable over the
lifetime of an experiment.
\end{abstract}


\maketitle

\section{Introduction}

One of the primary motivations in the original derivation of the
Gross-Pitaevskii equation was to describe vortices in
superfluids~\cite{pitaevskii1961,gross1961,fetter1972}. The
quantization of vorticity is a central difference between classical
fluids and superfluids~\cite{donnelly1991}.  The Gross-Pitaevskii
equation has proven to be an excellent model for weakly interacting,
dilute
atomic~\cite{anderson1995,davis1995,bradleyCC1995,bradleyCC1997,dalfovo1999,leggett2001}
and molecular~\cite{greiner2003,jochim2003b,zwierlein2003}
Bose-Einstein condensates (BEC's), which are generally superfluid.
In the context of optics the Gross-Pitaevskii equation is known as
the nonlinear Schr\"odinger equation (NLSE)~\cite{agrawal1995}.
Vortices~\cite{matthews1999,madison2000}, as well as their
one-dimensional analog, solitons~\cite{burger1999,denschlag2000},
have been observed in BEC's in numerous experiments.  The NLSE
describes these observations
well~\cite{ruprecht1995,fetter2001,feder1999,williams1999}.

However, there are in fact much richer vortex and soliton structures
yet to be observed.  One such structure is the \emph{ring
soliton}~\cite{theocharis2003,kivshar1994,dreischuh1996,frantzeskakis2000,nistazakis2001,neshev1997,dreischuh2002},
a soliton extended into two dimensions which loops back on itself to
form a ring, i.e., a radial node.  This appears as an axisymmetric
radial node of the condensate; set at the right distance from the
origin, it becomes a stationary state.  In the optics context, the
ring soliton is well known to be unstable to the snake instability,
whereby it decays into vortex-anti-vortex pairs. In this article, we
not only describe vortices and radially excited states of BEC's with
high precision, but we also show that in a harmonic trap the decay
time can be long compared to the classical oscillation period of the
trap and even the lifetime of the condensate itself~\cite{optics}.

We consider three cases for the external, or trapping potential
$V(\vec{r})$. First, $V(\vec{r})=0$ corresponds to an infinitely
extended condensate and leads to solutions of beautiful mathematical
form. Second, $V(\vec{r})=V(r)=0$ for $r\leq R$ and $V(r)=\infty$
for $r>R$, corresponds to a disk in two dimensions and a sphere in
three, i.e., an infinite well.  This introduces confinement into the
problem, connects heuristically with the first case and general
knowledge of solutions to Schr\"odinger equations, and serves as a
bridge to the experimental case of a harmonic trap. Third,
$V(\vec{r})=\frac{1}{2}M[\omega^2 (x^2+y^2)+\omega_z^2 z^2]$, with
$\omega \ll \omega_z$, where $M$ is the atomic mass and
$\omega,\omega_z$ are the classical oscillation frequencies,
corresponds to a highly oblate harmonic trap, which is most relevant
to present experiments.

This work follows in the spirit of a previous set of investigations
of the {\it one-dimensional} NLSE, for both repulsive and attractive
nonlinearity~\cite{carr2000a,carr2000b}. In that work it was
possible to obtain all stationary solutions in closed analytic form.
In the present cases of two and three dimensions, we are unaware of
an exhaustive class of closed form solutions but instead use a
combination of analytical and numerical techniques to elicit the
stationary and stability properties of similar solutions.  Here, we
treat the case of repulsive atomic interactions; as in our previous
work on one dimension, the attractive case has been treated
separately~\cite{carr2004k}, due to the very different character of
the solutions.

Equations similar to the NLSE are often used as models for classical
and quantum systems. Thus a tremendous amount of theoretical work
has been done on vortices, to which the reader is referred to Fetter
and Svidzinsky on BEC's~\cite{fetter2001}, Donnelly on Helium
II~\cite{donnelly1991}, and Saffman on classical
vortices~\cite{saffman1992} as good starting points for
investigations of the literature.  As NLSE-type equations apply in
many physical contexts, our results are widely applicable beyond the
BEC.

The article is outlined as follows.  The derivation of the
fundamental differential equations is presented in
Sec.~\ref{sec:equations}.  In Sec.~\ref{sec:vortices} the ground
state and vortices in two dimensions are presented.  In
Sec.~\ref{sec:rings} the stationary radial excitations of these
solutions are illustrated.  In Sec.~\ref{sec:spheres} the ground
state and its radial excitations in three dimensions are treated. In
Sec.~\ref{sec:stability} the stability properties of solution types
containing ring solitons are discussed. Finally, in
Sec.~\ref{sec:conclusions}, we discuss the results and conclude.

\section{Fundamental Equation}
\label{sec:equations}

The fundamental differential equation is derived as follows. The
NLSE, which models the mean field of a
BEC~\cite{pitaevskii1961,gross1961,dalfovo1999}, is written as \be
\label{eqn:gpe} \left[ -\frac{\hbar^2}{2M} \nabla^2 +g\,|\psi|^2 +
V(\vec{r})\right]\psi =i\hbar\frac{\partial}{\partial t}\psi \, ,
\ee where $V(\vec{r})$ is an external potential, $g\equiv
4\pi\hbar^2 a_s/M$, $a_s$ is the $s$-wave scattering length for
binary interaction between atoms with $a_s > 0$, since this is the
repulsive case, and $M$ is the atomic mass. The condensate order
parameter
$\psi=\psi(\vec{r},t)\equiv\sqrt{n(\vec{r},t)}\exp[iS(\vec{r},t)]$,
where $n(\vec{r},t)$ is the local atomic number density and
$\vec{v}(\vec{r},t)=(\hbar/M)\vec{\nabla} S(\vec{r},t)$ is the local
superfluid velocity.  Note that, in two dimensions, the coupling
constant $g$ is renormalized by a transverse
length~\cite{petrov2000,olshanii1998,petrov2000b,carr2000e}.

We assume cylindrical or spherical symmetry of both the external
potential and the order parameter in two or three dimensions,
respectively. This has the effect of reducing Eq.~(\ref{eqn:gpe}) to
one non-trivial spatial variable. Specifically, wavefunctions of the
form \be \psi(\vec{r})=f_m(r)\exp(i m \phi)\exp(-i \mu
t/\hbar)\exp(i\theta_0) \label{eqn:assume}\ee are treated, where $m$
is the winding number, $\mu$ is the eigenvalue, also called the
chemical potential, $\phi$ is the azimuthal coordinate, $r$ is the
radial coordinate in two or three dimensions, and $\theta_0$ is a
constant phase which may be taken to be zero without loss of
generality.

Assuming an axisymmetric stationary state in two dimensions of the
form given in Eq.~(\ref{eqn:assume}), Eq.~(\ref{eqn:gpe}) becomes
\begin{eqnarray} -\frac{\hbar^2}{2M}\left(\frac{\partial^2}{\partial r^2}
+\frac{1}{ r}\frac{\partial}{\partial r}-\frac{m^2}{ r^2} \right)
f_m\nonumber\\
+g f_m^3 +V(r) f_m -\mu f_m=0\,. \label{eqn:gpe2dunits}
\end{eqnarray} Assuming spherical symmetry in three dimensions,
one finds a similar equation,
\be -\frac{\hbar^2}{2M}\left(\frac{\partial^2}{\partial r^2}
+\frac{2}{ r}\frac{\partial}{\partial r} \right) f_0 +g f_0^3 +V(r)
f_0-\mu f_0=0\,. \label{eqn:gpe3dunits} \ee
In Eq.~(\ref{eqn:gpe3dunits}) it is assumed that $m=0$, in keeping
with the spherical symmetry. This is an important special case, as
it includes the ground state. In the remainder of this work, $m$
will be taken as non-negative since $f_{-|m|}=f_{|m|}$.

The physically relevant solutions of Eqs.~(\ref{eqn:gpe2dunits})
and~(\ref{eqn:gpe3dunits}) include the ground state, vortices, ring
solitons, and spherical shells, as we will show in the following
three sections.  A variety of solution methods are used to treat the
wavefunction in different regions of $r$, as discussed in
Appendix~\ref{app:numerics}.  Different rescalings of
Eqs.~(\ref{eqn:gpe2dunits})-(\ref{eqn:gpe3dunits}) are appropriate
for the three potentials we consider: constant, infinite well, and
oblate harmonic.  These are treated briefly in the following
subsections.

\subsection{Constant Potential}

A constant potential has no direct experimental realization.
However, it does have mathematical properties which are helpful in
understanding the confined case, not to mention beautiful in
themselves.  For instance, the vortex solution manifests as a
boundary between divergent and non-divergent solutions, as shall be
explained.

The potential $V(r)=V_0$ can be taken to be zero without loss of
generality. Then the variables can be rescaled as
\bea \eta_m&\equiv&\sqrt{\frac{g}{\mu}}f_m\, ,\label{eqn:defeta}\\
\chi&\equiv&\frac{\sqrt{2M\mu}\,r}{\hbar}\, .\label{eqn:defchi}\eea
Note that the radial coordinate is scaled to the length associated
with the chemical potential.  Then Eq.~(\ref{eqn:gpe2dunits})
becomes \be \eta_m''+\frac{1}{\chi}\eta_m'-\frac{m^2}{\chi^2}\eta_m
- \eta_m^3 +\eta_m=0 \, ,\label{eqn:gpe2d}\ee and
Eq.~(\ref{eqn:gpe3dunits}) becomes \be
\eta_0''+\frac{2}{\chi}\eta_0'-\eta_0^3 +\eta_0=0 \,
,\label{eqn:gpe3d}\ee where $\mu$ has been assumed to be positive,
since this is the physically meaningful case for repulsive atomic
interactions.  Note that, in these units, the length scale of a
vortex core is on the order of unity.

\subsection{Infinite Potential Well in Two and Three Dimensions}

Let us first consider the two dimensional infinite well.  A confined
condensate may be obtained by placing an infinite potential wall at
fixed $r=\sqrt{x^2+y^2}$, at any node of the wavefunction
$\psi(r,\phi)$. This treats either a condensate tightly confined in
$z$ or a cylinder of infinite $z$ extent. In either case, one
derives a 2D NLSE from the 3D one by projecting the $z$ degree of
freedom onto the ground state and integrating over it. This leads to
a straightforward renormalization of the coefficient of the
nonlinear, cubic term.  Extremely high potential wells in the
$x$-$y$ plane have been created in BEC experiments via higher order
Gauss-Laguerre modes of optical traps~\cite{bongs2001}.

For the purposes of our numerical algorithm outlined in
Appendix~\ref{app:numerics}, it is convenient to keep the same
scalings as Eqs.~(\ref{eqn:defeta})-(\ref{eqn:defchi}).  Then the
normalization has to be treated with some care. The normalization of
$\psi$ in two dimensions for a BEC of $N$ atoms in an infinite
cylindrical well of radius $R$ is given by \be \int_0^{2\pi}
d\phi\int_0^R dr\, r |\psi(r,\phi;t)|^2=N \, .\ee After the change
of units given by Eqs.~(\ref{eqn:defeta}) and~(\ref{eqn:defchi}) the
normalization becomes \be \int_0^{\chi_R} d\chi\,\chi
[\eta_m(\chi)]^2 = 2 \mathcal{N}\,, \label{eqn:norm}\ee where \bea
\mathcal{N}\equiv
\frac{M}{2\pi\hbar^2}gN\label{eqn:effnonlin}\,,\\
\chi_R\equiv \sqrt{\frac{2M\mu}{\hbar^2}}R \label{eqn:efflength}\eea
are the effective nonlinearity and cylinder radius.

The three-dimensional infinite well is treated similarly, with the
normalization being $\mathcal{N}$, rather than $2\mathcal{N}$, due
to the extra angular integration.

The details of the algorithm for calculation of quantized modes in
two and three dimensions is discussed in
Appendix~\ref{ssec:infinitewell}.

\subsection{Oblate Harmonic Trap}
\label{ssec:oblate}

For the harmonic potential we shall focus on the case of a highly
oblate trap, which is the experimentally relevant one to obtain
axial symmetry in two effective dimensions~\cite{ketterle1999}.
Then, again projecting onto the ground state in $z$, the rescaled
nonlinearity $\mathcal{N}$ has a simple interpretation: \be
\mathcal{N}= 2 a_s N \sqrt{M\omega_z/ 2\pi \hbar}\,,\ee where
$\omega_z$ is the angular frequency of the trap in the $z$
direction. The trap is isotropic in the remaining two directions,
$\omega\equiv \omega_x=\omega_y$. All energies can then be scaled to
$\hbar \omega$, lengths to $\ell\equiv \sqrt{\hbar/M\omega}$, etc.,
as follows: \be
-\frac{1}{2}\left(\tilde{f}_m''+\frac{1}{\tilde{r}}\tilde{f}_m'-\frac{m^2}
{\tilde{r}^2}\tilde{f}_m\right) + \tilde{f}_m^3
+\frac{1}{2}\tilde{r}^2\tilde{f}_m-\tilde{\mu}\tilde{f}_m=0 \,
,\label{eqn:gpeHarmonic}\ee where the tildes refer to harmonic
oscillator scalings.  Explicitly, $\tilde{r}\equiv r/\ell$,
$\tilde{f}_m\equiv \ell f_m$, $\tilde{\mu}\equiv\mu/\hbar\omega$,
and the normalization is \begin{equation} \int d\tilde{r}\,
\tilde{r}
\,|\tilde{f}_m|^2=\mathcal{N}\,.\label{eqn:honorm}\end{equation} The
main difference between Eqs.~(\ref{eqn:gpeHarmonic}) and
Eq.~(\ref{eqn:gpe2d}) is that an extra parameter must be set in the
numerical algorithm of Appendix~\ref{app:numerics}, i.e., the
rescaled chemical potential. Then the normalization is determined
from Eq.~(\ref{eqn:honorm}).

\section{Ground State and Vortices}
\label{sec:vortices}

\begin{figure}[t]
\centerline{\epsfysize=15cm \leavevmode  \epsfbox{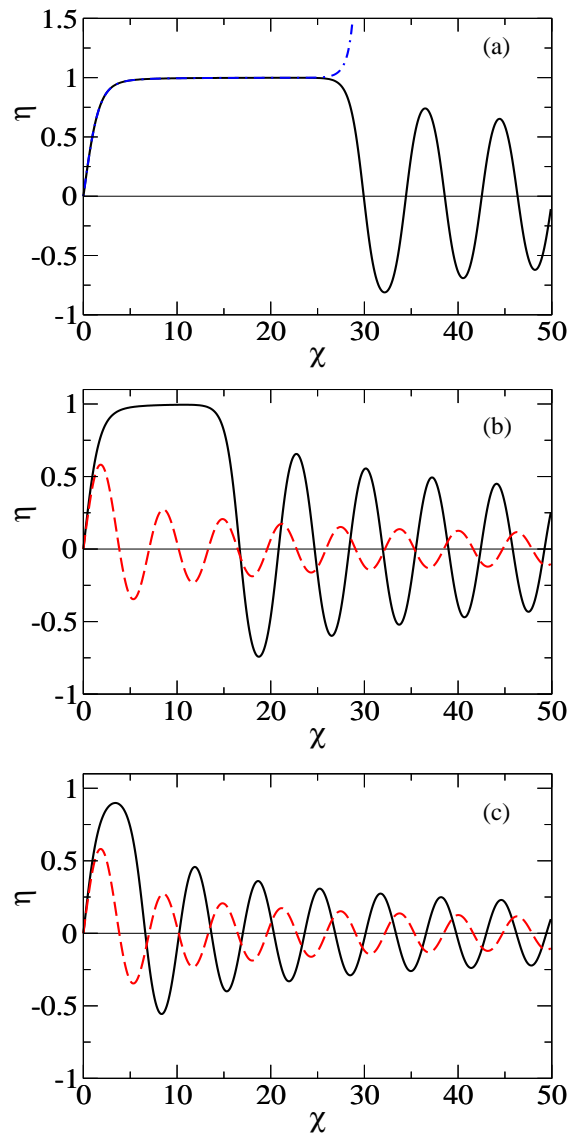}}
\caption{\label{fig:1} (color online) {\it Approaching the vortex
solution.} Shown is the dependence of the wavefunction on the
precision in the critical determining coefficient $a_0^{(m=1)}$ for
a quantum vortex stationary state of the nonlinear Schr\"odinger
equation in two dimensions. (a) The vortex solution forms a boundary
between convergent ($a_0=a_0^{(1)}(1-10^{-16})$, solid black curve)
and divergent ($a_0=a_0^{(1)}(1+10^{-16})$, dot-dashed blue curve)
solutions.  As the precision is reduced, the first node moves
towards the origin and the solution approaches the Bessel function:
(b) 8 digits of precision; (c) 2 digits of precision. In (b) and (c)
the regular Bessel function solution to the {\it linear}
Schr\"odinger equation is shown for comparison~\cite{abramowitz1964}
(dashed red curve).  Note that all axes are dimensionless.}
\end{figure}

In order to solve Eqs.~(\ref{eqn:gpe2dunits})
and~(\ref{eqn:gpe3dunits}), we use a numerical shooting method, as
discussed in App.~\ref{ssec:shooting}.  Two initial conditions are
required, as the NLSE is second order.  These are provided by a
Taylor expansion around $\chi=0$, as described in
Sec.~\ref{ssec:power}.  By this method a single parameter is
sufficient to determine the solution.  This parameter, $a_0$, is the
lowest non-zero coefficient in the Taylor series.  High precision is
required, as discussed in the appendix, with the number of digits of
precision determining where the numerical method either converges or
diverges.

For the first potential we consider, $V(\chi)=0$, the ground state,
which is obtained for $m=0$, lies precisely on the border between
convergence and divergence of the algorithm. The value of $a_0$
which is exactly on this border we term $a_0^{(0)}$. The
wavefunction $\eta_0(\chi)$ is constant and, according to our
scalings, is simply $\pm\sqrt{\mu/g}=\pm 1$. Then $a_0^{(0)}=\pm 1$
and the precision is infinite.  Values of $|a_0|$ larger than 1 lead
to a divergent solution, while values of $|a_0|$ less than 1 lead to
a convergent one. In the latter case the wavefunction oscillates an
infinite number of times and approaches zero, as will be discussed
in Sec.~\ref{sec:rings}. Thus, in general, $|\eta_0|$ can approach
only three asymptotic values: 0, 1, and $\infty$.

For the case of non-zero winding number, one finds a central vortex.
As in the case of the ground state,
$\lim_{\chi\rightarrow\infty}|\eta_m|^2\equiv\bar{n}=\mu/g$ is the
asymptotic density of the vortex state.  In this region the spatial
derivatives yield zero and $\eta_m(\chi)\rightarrow \pm 1$ as
$\chi\rightarrow\infty$.  The vortex again lies on the border
between divergence and convergence of our algorithm, given by a
single parameter $a_0^{(m)}$ which determines the whole Taylor
series. In Appendix~\ref{app:numerics} the precision issues are
discussed in detail.  In Fig.~\ref{fig:1} is illustrated the
algorithmic approach to the vortex solution for zero external
potential.  The effect of the number of digits of precision is shown
in detail, and further explained in Appendix~\ref{app:numerics},
where the best values of $a_0^{(m)}$ for winding numbers $m=1$ to
$m=5$ are given.

For the second potential, an infinite well in two dimensions, the
properly normalized ground state and vortex states are produced by
the simple prescription give in Appendix~\ref{ssec:infinitewell}. An
example is shown for $m=1$ in Fig.~\ref{fig:2}(a).  The ground state
for three dimensions can be found by a similar method, and is shown
for $m=0$ in Fig.~\ref{fig:6}(a).

For the third potential, a strongly oblate harmonic trap, the form
and stability of both the ground state and vortex solutions in a
harmonic trap have already been thoroughly studied
elsewhere~\cite{ruprecht1995,dodd1996a,pu1999}. Our algorithm
reproduces all the relevant results of these authors from the
non-interacting to the Thomas-Fermi regime, as we explicitly
verified in detail in the case of Ref.~\cite{pu1999}.

\section{Ring Solitons}
\label{sec:rings}

Ring solitons can be placed concentrically to form a stationary
state.  In an extended system a denumerably infinite number are
required, as already indicated in Fig.~\ref{fig:1} and discussed in
further detail in Sec.~\ref{ssec:cylindricalasymptotic} below.  For
an infinite well in two dimensions, radially quantized modes are
distinguished by the number of concentric ring solitons, as
discussed in Sec.~\ref{ssec:cylindricalbox} and illustrated in
Fig.~\ref{fig:2}.  Note that, for \emph{attractive} nonlinearity,
the number of rings can vary from one to infinity, even for an
infinitely extended condensate, as discussed in
Ref.~\cite{carr2004k}.

\subsection{Quantized Modes in the Cylindrical Infinite Well}
\label{ssec:cylindricalbox}

\begin{figure}[t]
\centerline{\epsfxsize=8cm \leavevmode  \epsfbox{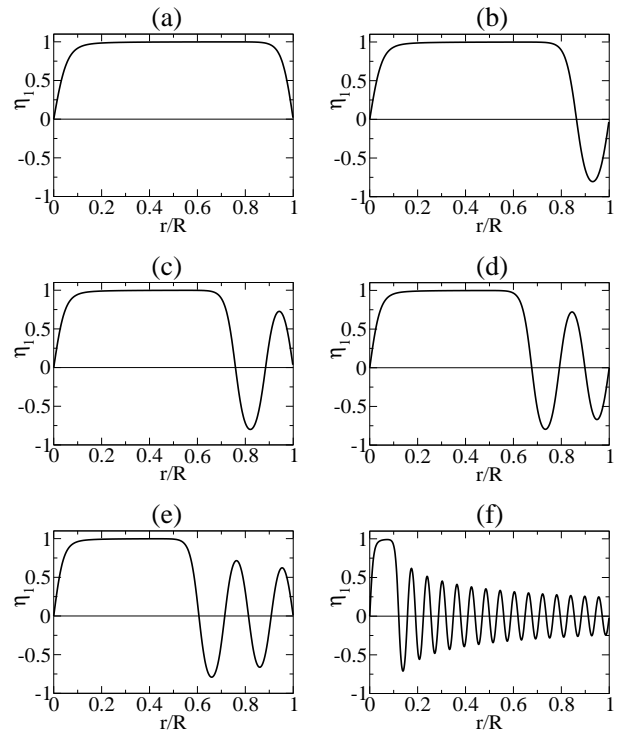}}
\caption{\label{fig:2} {\it Ring solitons in an infinite well}.
Shown is the form of the wavefunction $\eta_1$ for (a) the ground
state, (b)-(e) the first four axisymmetric excited states, and (f) a
highly excited state, all as a function of the radial coordinate $r$
scaled to the cylinder size $R$. Cylindrical box boundary conditions
in cylindrical polar coordinates are assumed. The central vortex has
winding number $m=1$.  The quantized modes (b)-(f) correspond to
increasing numbers of concentric ring solitons.  Here, the case of
strong nonlinearity is illustrated, with $2\mathcal{N}=402$. Note
that all axes are dimensionless.}
\end{figure}

\begin{figure}
\centerline{\epsfxsize=8cm \epsfysize=15cm \leavevmode
\epsfbox{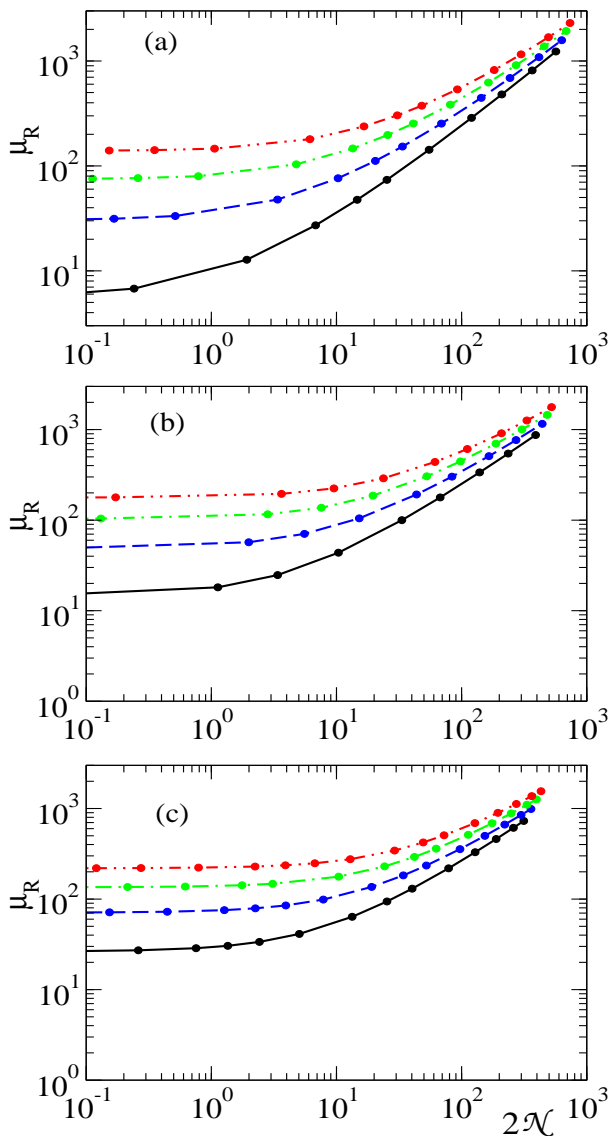}} \caption{\label{fig:3} (color online) {\it
Ring solitons in an infinite well}.  Shown are the eigenvalue
spectra $\mu_R$ as a function of the normalization $2\mathcal{N}$.
The winding numbers (a) $m=0$, (b) $m=1$, and (c) $m=2$ are
illustrated for the ground state (solid black curve), and the first
three excited states (blue dashed, green dot-dashed, and red
double-dot-dashed curves). The circles show the actual data points.
Note that all axes are dimensionless as indicated in the text, and
on the same log-log scale.}
\end{figure}

In Fig.~\ref{fig:2} are shown the wavefunctions for the ground state
and first three excited states for a fixed strong nonlinearity
$\mathcal{N}$ in the presence of a singly quantized central vortex,
i.e., winding number $m=1$. The weaker the nonlinearity and the
larger the number of nodes in the solution, the closer it resembles
the regular Bessel function $J_m(\chi)$. In an appropriately scaled
finite system, the relative weight of the kinetic term to the mean
field term in the NLSE increases strongly with the number of nodes.
We note the contrast with the distribution of nodes for the
corresponding 1D NLSE~\cite{carr2000a,carr2000b}, where the nodes
are evenly spaced even in the extremely nonlinear limit.  The value
of the nonlinearity was chosen to be $2\mathcal{N}=402$. For
transverse harmonic confinement of angular frequency
$\omega_z=2\pi\times 100$ Hz and $^{87}$Rb, which has a scattering
length of $a_s=5.77$ nm, this corresponds to $N\simeq 10,000$ atoms.
Note that, since $\eta$ is scaled to $g/\mu$, and $\mu$ depends on
the number of nodes, the vertical scaling is different in each panel
of Fig.~\ref{fig:2}.

The eigenvalue spectra for the ground state and the first two
excited states are shown in Fig.~\ref{fig:3}, with $\mu$ scaled to
the radius of the infinite well, \begin{equation} \mu_R\equiv
\frac{2MR^2}{\hbar^2}\mu\,.\end{equation} The winding numbers $m=0$,
$m=1$, and $m=2$ are illustrated on a log-log scale. Clearly there
are two regimes. For small $\mathcal{N}$, $\mu_R$ is independent of
the norm.  This must be the case near the
linear-Schr\"odinger-equation regime, since $\mu_R$ must approach
the eigenvalues of the regular Bessel function $J_m(\chi)$ which
solves Eq.~(\ref{eqn:gpe2d}) with no cubic term. One finds
$(\mu_{R})_j\rightarrow (\chi_j^{\mathrm{linear}})^2$, where
$\chi_j^{\mathrm{linear}}$ is the known value of the
$j^{\mathrm{th}}$ node of the Bessel function~\cite{abramowitz1964}
and $j$ also refers to the $j^{\mathrm{th}}$ quantized mode.

For large $\mathcal{N}$, the figure shows that $\mu_R\propto
\mathcal{N}$. This dependence can be understood analytically in the
case of the Thomas-Fermi-like profile~\cite{dalfovo1999} for the
lowest energy state with winding number $m=1$. Consider the scaling
\bea\rho&\equiv&
r/R\,\\
\kappa(\rho)&\equiv& \frac{1}{\sqrt{N}}f_m(r/R)\,.\eea  Then
Eq.~(\ref{eqn:gpe2dunits}) becomes \be
\frac{\partial^2\kappa}{\partial\rho^2}+\frac{1}{\rho}
\frac{\partial\kappa}{\partial\rho}-\frac{m^2}{\rho^2}\kappa -4\pi
R^2\mathcal{N}\kappa^3+\mu_R\kappa=0\,.\ee The Thomas-Fermi profile
is obtained by dropping the derivatives: \be
\kappa_{\mathrm{TF}}(\rho)=\sqrt{\frac{\mu_R}{4\pi R^2\mathcal{N}}
\left(1-\frac{\rho_m^2}{\rho^2} \right)}\,\ee where \be \rho_m\equiv
\frac{m}{\sqrt{\mu_R}}\ee is the core size and
$\kappa_{\mathrm{TF}}$ is zero for $\rho\leq \rho_m$.  The
normalization condition is \be 2\pi R^2 \int_{\rho_m}^1 d\rho\,\rho
\left[\kappa_{\mathrm{TF}}(\rho)\right]^2=1\, .\ee Then the chemical
potential in units of the energy associated with the cylinder radius
is \be \mu_R = \frac{4\mathcal{N}}{1-\rho_m^2 +\rho_m^2\ln
(\rho_m^2)}\, .\ee The limit $\rho_m\ll 1$ is consistent with the
Thomas-Fermi-like profile, which neglects the radial kinetic energy.
In this limit, one finds \be \mu_R \simeq
4\mathcal{N}\left[1+\rho_m^2-\rho_m^2\ln
\left(\rho_m^2\right)\right]\,.\ee For $\rho_m^2\leq 0.03$, as is
the case in the right hand side of Fig.~\ref{fig:3}(a)-(c) where
$\mu_r\simeq 10^3$, the dependence on $\rho_m^2$ is a less than 1\%
perturbation.

\subsection{Quantized Modes in the Strongly Oblate Harmonic Trap}
\label{ssec:harmonic}

\begin{figure}
\centerline{\epsfxsize=8cm \leavevmode  \epsfbox{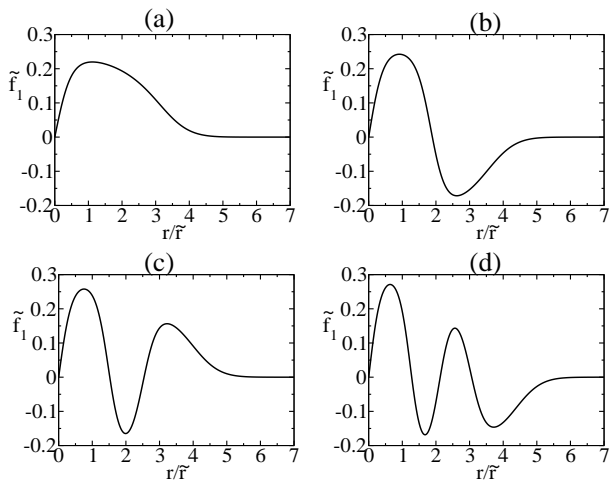}}
\caption{\label{fig:4} {\it Ring solitons in a harmonic trap}. Shown
is the form of the wavefunction $\tilde{f}_1$ for (a) the ground
state and (b)-(d) the first three axisymmetric excited states, all
as a function of the radial coordinate $\tilde{r}$.  The harmonic
trap is strongly oblate, so that it is effectively two-dimensional.
The central vortex has winding number $m=1$. The quantized modes
(b)-(d) correspond to increasing numbers of concentric ring
solitons.  Here, the case of strong nonlinearity is illustrated for
(a), with $\mathcal{N}=100$; for three rings, i.e., (d), the
solution already appears nearly linear. Note that the tildes signify
that all axes are in harmonic oscillator units.}
\end{figure}

\begin{figure}
\centerline{\epsfxsize=8cm \epsfysize=15cm \leavevmode
\epsfbox{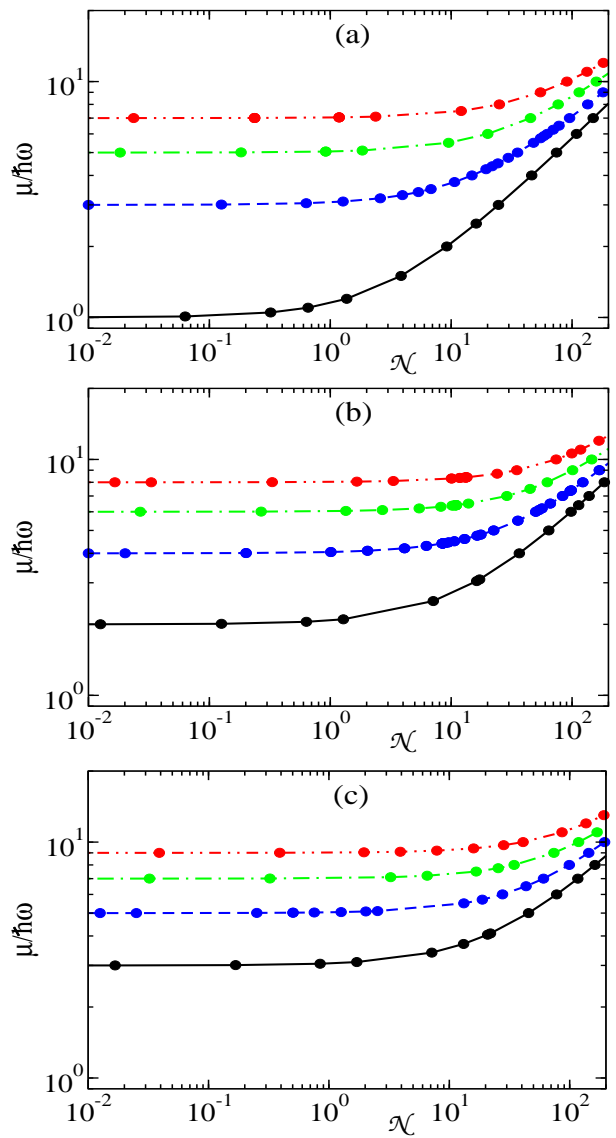}} \caption{\label{fig:5} {\it Ring solitons
in a harmonic trap}. Shown are the eigenvalue spectra $\mu$ as a
function of the normalization $\mathcal{N}$, all in harmonic
oscillator units. The ground state (solid black curve) and the first
three isotropic excited states (blue dashed, green dot-dashed, and
red double-dot-dashed curves) are illustrated for a winding number
of (a) $m=0$, (b) $m=1$ and (c) $m=2$.  The circles show the actual
data points. Note that all axes are dimensionless and on a log-log
scale.}
\end{figure}

In figure~\ref{fig:4} is shown the wavefunction for fixed strong
nonlinearity for the ground state and first three excited states. As
in Sec.~\ref{ssec:cylindricalbox}, adding nodes to the wavefunction
drives the system towards the linear regime, where the solution is
Bessel-function-like.  In Fig.~\ref{fig:5} are shown the spectra for
the ground state and first three excited states, for a winding
number of $m=0,1,2$.  Note that the chemical potential is rescaled
to the harmonic oscillator energy.  As in
Sec.~\ref{ssec:cylindricalbox}, there are two regimes. For large
nonlinearity a Thomas-Fermi approximation can be applied to obtain
the asymptotic dependence of the chemical potential on the
nonlinearity, similar to the procedure of
Sec.~\ref{ssec:cylindricalbox}.

\subsection{Asymptotic Behavior for Zero External Potential}
\label{ssec:cylindricalasymptotic}

For $\chi\rightarrow\infty$ and $|a_0|< |a_0^{(m)}|$ the
wavefunction approaches zero in an infinitely extended system.  Thus
one expects that the nonlinear term $\eta^3$ in
Eq.~(\ref{eqn:gpe2d}) is negligible in comparison to the other
terms, and the differential equation returns to the usual defining
equation for the Bessel functions. The asymptotic form of the
regular Bessel function $J_m(\chi)$ is~\cite{abramowitz1964} \be
J_m(\chi)=\sqrt{\frac{2}{\pi\chi}}\cos
\left[\chi-\frac{m\pi}{2}-\frac{\pi}{4}\right]
\,,\label{eqn:besselexpand}\ee to leading order in the amplitude and
the phase.  However, one cannot neglect the effect of the cubic term
on the phase shift, as may be seen by the following considerations.

The asymptotic form of the Bessel function can be derived via the
semiclassical WKB approximation~\cite{brack1997}.  The phase shift
of $\pi/4$ can be derived by analytical continuation or other
means~\cite{landau1977}.  The semiclassical requirement that the
de Broglie wavelength be small compared to the length scale of the
change in potential is not quite satisfied near the origin.  The
rescaling $y=y_0\ln(\chi)$ suffices to map the problem onto the
usual semiclassical one.  One can avoid the rescaling by the
substitution $m^2\rightarrow m^2 -\frac{1}{4}$.  Then the term
$m\pi/2$ in the phase shift follows
directly~\cite{langer1937,berry1972}.  We use this simpler method
in order to derive the phase shift in the nonlinear problem.

The semiclassical momentum is \be
p(\chi)\equiv\sqrt{1-V_{\mathrm{eff}}(\chi)}\,
,\label{eqn:momentum1}\ee where the effective potential is \be
V_{\mathrm{eff}}(\chi)=\frac{m^2}{\chi^2}+[\eta_m(\chi)]^2\,
.\label{eqn:effpot}\ee Taking the nonlinear term as perturbative, to
lowest order Eq.~(\ref{eqn:effpot}) becomes \bea
V_{\mathrm{eff}}(\chi)&\simeq&\frac{m^2}{\chi^2}
+\frac{B^2}{\chi}\cos^2\left(\theta_{m}\right)\, ,\label{eqn:effpot2}\\
\theta_{m}&\equiv&\chi-\frac{m\pi}{2}-\frac{\pi}{4}\,,\eea where $B$
is a constant coefficient of the amplitude of the wavefunction. In
the linear case, it is conventionally taken as $B=\sqrt{2/\pi}$.
Expanding Eq.~(\ref{eqn:momentum1}) for large $\chi$, one finds \be
p(\chi)\simeq 1- \frac{B^2\cos^2\theta_{m}}{2\chi}
-\frac{m^2+\frac{1}{8}B^4\cos^4\theta_m}{2\chi^2}
\label{eqn:momentum2}\,.\ee

The semiclassical form of the wavefunction
is~\cite{brack1997,landau1977} \be \eta_m \simeq
\frac{B}{\sqrt{\chi}}\cos\left(S-\frac{\pi}{4}\right) \ee to
leading order, where \be S\equiv\int_0^{\chi} d\chi'\,p(\chi')\,
\label{eqn:action}\ee is the semiclassical action. Upon
substitution of Eq.~(\ref{eqn:momentum2}) one finds the form of
the wavefunction to leading order in the amplitude and the phase,
\bea \eta_m\simeq \frac{B}{\sqrt{\chi}}\cos\left[
\chi-\frac{B^2}{4}\ln(\chi)
\right. \nonumber\\
\left.-\frac{m\pi}{2}-\frac{\pi}{4}+\delta(a_0,m)\right]\,,
\label{eqn:asymptotic}\eea where $\delta$ is a phase shift which
depends on the determining coefficient $a_0$ and the winding number
$m$.  This coefficient cannot be analytically determined by
Eq.~(\ref{eqn:action}) since the large $\chi$ form of the
wavefunction was used, while the phase shift is due to its behavior
in the small $\chi$ region. The amplitude coefficient $B$ is a free
parameter, the square of which is related to the mean number
density.

The form of Eq.~(\ref{eqn:asymptotic}) resembles that of the
Coulomb function~\cite{abramowitz1964}, in that it has a
$\ln(\chi)$ dependence in the phase.  This is due to the $1/\chi$
term in the effective potential in Eq.~(\ref{eqn:effpot2}).  It is
in this sense that the nonlinear term in Eq.~(\ref{eqn:gpe2d})
cannot be neglected, even as $\eta\rightarrow 0$.

\section{Spherical Shells}
\label{sec:spheres}

Spherical shells are the three-dimensional analog of ring solitons.
Quantized modes in the confined case involve successive numbers of
nested nodal spherical shells.  In this section, we will treat the
two cases of zero potential and an infinite spherical well.  A power
series solution of Eq.~(\ref{eqn:gpe3d}) may be developed by
substitution of Eq.~(\ref{eqn:taylor}).  This leads to a solution
similar to that of Sec.~\ref{ssec:power}.  All coefficients in the
power series are given as polynomials in the determining coefficient
$a_0$, which is the value of the wavefunction at the origin. The
special solution $\eta_0(\chi)=a_0=\pm 1$ is the ground state in an
extended system. Thus $a_0^{(0)}=\pm 1$. Positive values of $a_0$
which are larger than unity lead to a divergent solution. Those less
than unity lead to a convergent solution which approaches zero as
$\chi\rightarrow \infty$.

\subsection{Quantized Modes in the Spherical Infinite Well}
\label{ssec:sphericalbox}

\begin{figure}
\centerline{\epsfxsize=8cm \leavevmode  \epsfbox{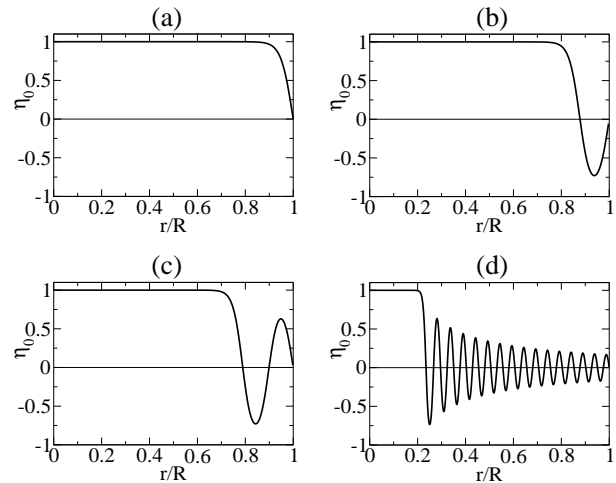}}
\caption{\label{fig:6} {\it Spherical shell solutions}.  Shown is
the form of the wavefunction for (a) the ground state, (b)-(c) the
first two excited isotropic states, and (d) a highly excited
isotropic state.  The boundary conditions are an infinite spherical
well in three dimensions.  The winding number is zero.  The case of
strong nonlinearity is illustrated, with $\mathcal{N}=251$. Note
that all axes are dimensionless.}
\end{figure}

\begin{figure}
\centerline{\epsfxsize=8cm \leavevmode  \epsfbox{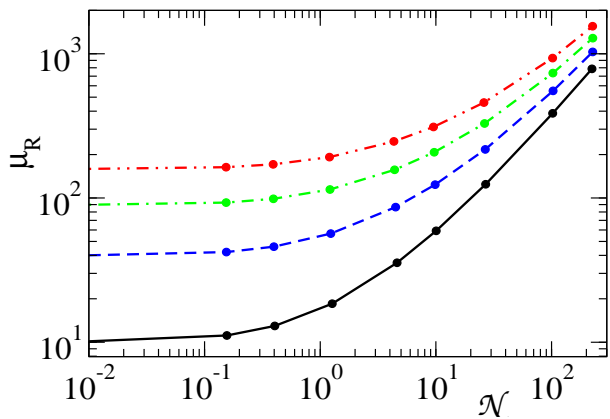}}
\caption{\label{fig:7} {\it Spherical shell solutions}.  Shown are
the eigenvalue spectra $\mu_R\equiv(2MR^2/\hbar^2)\mu$ as a function
of the normalization $\mathcal{N}$. The ground state (solid black
curve) and the first three isotropic excited states (blue dashed,
green dot-dashed, and red double-dot-dashed curves) are illustrated
for a winding number of $m=0$.  The circles show the actual data
points.  Note that all axes are dimensionless and on a log-log
scale.}
\end{figure}

Solutions can be quantized in the three-dimensional spherical well
in the same way as the two-dimensional case.  The solution methods
are identical to those of Sec.~\ref{ssec:cylindricalbox}. In
Fig.~\ref{fig:6} are shown the ground state, the first and second
excited states, and a highly excited state.  A fixed nonlinearity of
$\mathcal{N}=251$ was chosen.  In Fig.~\ref{fig:7} is shown the
eigenvalue spectra on a log-log scale.  As in the two-dimensional
case of Fig.~\ref{fig:3}, there are two regimes. For small
$\mathcal{N}$, the eigenvalues are independent of the nonlinearity,
$(\mu_{R})_j\rightarrow (\chi_j^{\mathrm{linear}})^2$. The constant
$\chi_j^{\mathrm{linear}}$ is the distance to the $j$th nodes of the
spherical Bessel function $j_0(\chi)$ which solves
Eq.~(\ref{eqn:gpe3d}) with no cubic term~\cite{abramowitz1964}. For
large $\mathcal{N}$, one again finds a linear dependence. A simple
estimate based on the Thomas-Fermi profile for $m=0$, which is just
$f_0(r)=\mu/g$ for $r\leq R$ and zero otherwise, gives the chemical
potential of the ground state as $\mu_R = \mathcal{N}/3$.

\subsection{Asymptotic Behavior}
\label{ssec:sphericalasymptotic}

As $r\rightarrow\infty$ the spherical shell wavefunction approaches
zero in an infinitely extended system.  Just as in
Sec.~\ref{ssec:cylindricalasymptotic}, one can use the WKB
semiclassical approximation method to determine the asymptotic form
of the wavefunction.  The solution to Eq.~(\ref{eqn:gpe3d}) without
the cubic term is the spherical Bessel
function~\cite{abramowitz1964} \be
j_0(\chi)=\frac{\sin\chi}{\chi}\,,\ee where we have assumed the
wavefunction to be finite at the origin.  One can take the nonlinear
term as perturbative since, for sufficiently large $\chi$, the
linear form of the wavefunction must dominate. Then the effective
potential in the WKB formalism is \be V_{\mathrm{eff}}(\chi)\simeq
B^2\frac{\sin^2\chi}{\chi^2} \,,\ee where $B$ is the amplitude of
the wavefunction. From Eqs.~(\ref{eqn:momentum1})
and~(\ref{eqn:action}), the WKB action is \be S\simeq \chi -
\frac{B^2}{4}\frac{1}{\chi}+\delta'(a_0,m)\ee for large $\chi$,
where the phase shift $\delta'(a_0,m)$ cannot be determined from the
large $\chi$ behavior of the wavefunction. Then the asymptotic form
of the wavefunction is \be \eta_0(\chi)\rightarrow
\frac{B}{\chi}\sin\left[\chi-\frac{B^2}{4\chi}+\delta'(a_0,m)\right]\,.\ee

\section{Stability Properties}
\label{sec:stability}

\begin{figure}
\centerline{\epsfxsize=8cm \leavevmode \epsfbox{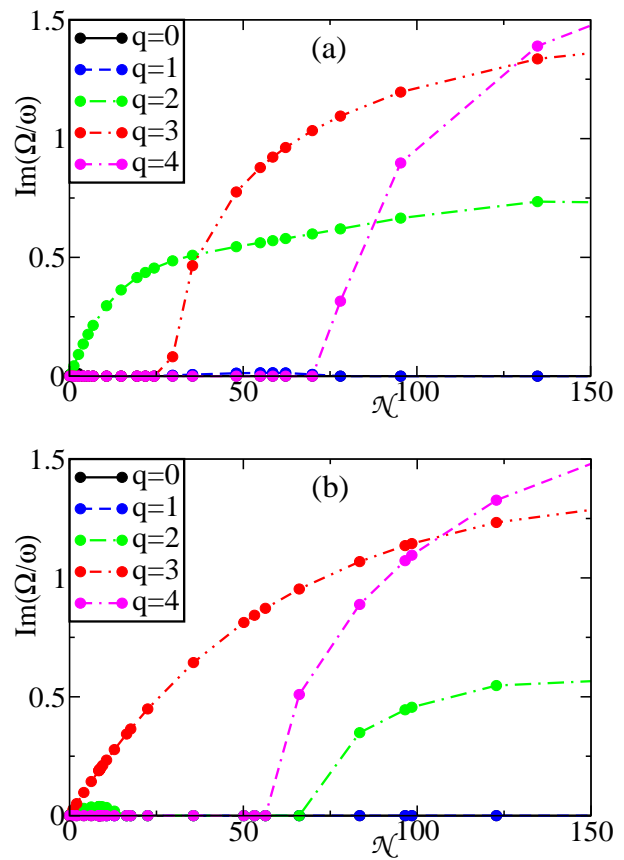}}
\caption{\label{fig:8} (color online) {\it Stability Properties of
Ring Solitons.}  Shown is the Boguliubov linear excitation frequency
as a function of the nonlinearity $\mathcal{N}$, all in harmonic
oscillator units. The index $q$ signifies the winding number of the
Boguliubov mode. (a) A ring soliton with no central vortex; for
$\mathcal{N}\lesssim 25$ the primary instability mode is $q=2$. (b)
A ring soliton in the presence of a central vortex of winding number
$m=1$; for $\mathcal{N}\lesssim 100$ the dominant instability is
$q=3$. The solutions are always formally unstable, though
instability times can be much longer than experimental timescales
for small nonlinearities.  The circles show the actual data points.}
\end{figure}

The stability properties of ring solitons in a strongly oblate
harmonic trap are of particular importance, as such solutions may be
realized in present experiments~\cite{theocharis2003}.  We perform
linear stability analysis via the well-known Boguliubov de Gennes
equations~\cite{svidzinsky1998,fetter2001}:
\begin{eqnarray}
{\cal L}u_j-g\psi^2v_j&=&\hbar\Omega_j u_j,\label{eqn:bog1}\\
{\cal L}v_j-g\left(\psi^*\right)^2u_j&=&- \hbar\Omega_j v_j,
\label{eqn:bog2}
\end{eqnarray}
where \be {\cal L} \equiv -\frac{\hbar^2}{2M}\nabla^2+V(\vec{r}) +
2g|\psi|^2-\mu\,,\ee and $\Omega_j$ is the eigenvalue.  In
Eqs.~(\ref{eqn:bog1})-(\ref{eqn:bog2}) $u_j$ and $v_j$
 are a complete set of
coupled quasiparticle amplitudes
 that obey the normalization condition \be \int
d^Dr\,\left(|u_j|^2-|v_j|^2\right) = 1\,,\ee with $D$ the number of
effective dimensions.  These amplitudes represent excitations
orthogonal to the condensate $\psi$.  They have a straightforward
quantum mechanical interpretation in terms of a canonical
transformation of the second-quantized Hamiltonian for binary
interactions via a contact potential of strength
$g$~\cite{fetter2001}.  Quasiparticles are superpositions of
particles (creation operators) and holes (annihilation operators).
Classically, they can be interpreted simply as linear perturbations
to the condensate.

If the eigenvalue $\Omega_j$ is real, the solution $\psi$ is stable.
If $\Omega_j$ has an imaginary part, then $\psi$ is unstable.  There
are significant subtleties in Boguliubov analysis; see the appendix
of Ref.~\cite{garay2001} for an excellent discussion.  In the
present effectively 2D potential with a condensate solution of the
form given in Eq.~(\ref{eqn:assume}), it is useful to redefine the
Boguliubov amplitudes as suggested by Svidzinsky and
Fetter~\cite{svidzinsky1998}:
\begin{equation}
\pmatrix{u( \vec{r}) \cr v(\vec{r})} =
\frac{e^{iq\phi}}{\ell}\,\pmatrix{e^{im\phi}\,\tilde
u_q(\tilde{r})\cr e^{-im\phi}\,\tilde
v_q(\tilde{r})}\,,\label{eqn:uvq}
\end{equation}
where $\tilde{r}=\sqrt{x^2+y^2}/\ell$, $\ell$ is the harmonic
oscillator length discussed in Sec.~\ref{ssec:oblate}, and we will
neglect perturbations in the $z$ direction, due to the strongly
oblate trap. Equation~(\ref{eqn:uvq}) represents a partial wave of
angular momentum $q$ relative to the condensate. Then, in harmonic
oscillator units (see Sec.~\ref{ssec:oblate}),
Eqs.~(\ref{eqn:bog1})-(\ref{eqn:bog2}) become
\begin{eqnarray}
\mathcal{L}_{+}\tilde{u}_q-|\tilde{f}_m|^2\tilde{v}_q=\frac{\Omega}{\omega}\tilde{u}_q,
\label{eqn:bog3}\\
\mathcal{L}_{-}\tilde{v}_q-|\tilde{f}_m|^2\tilde{u}_q=-\frac{\Omega}{\omega}\tilde{v}_q,
\label{eqn:bog4}
\end{eqnarray}
where
\begin{eqnarray}
\mathcal{L}_{\pm} &\equiv&
-\frac{1}{2}\left(\frac{\partial^2}{\partial
\tilde{r}^2}+\frac{1}{\tilde{r}}\frac{\partial}{\partial \tilde{r}}
- \frac{(q\pm
m)^2}{\tilde{r}^2}-\tilde{r}^2\right)\nonumber\\
&&+2|\tilde{f}_m|^2-\tilde{\mu}
\end{eqnarray}

The different centrifugal barriers inherent in $\mathcal{L}_{\pm}$
show that the two amplitudes behave differently near the origin,
with $\tilde u_q \propto \tilde{r}^{|q+m|} $ and $\tilde v_q\propto
\tilde{r}^{|q-m|}$ as $\tilde{r}\to 0$.  Note that the nonlinear
coefficient is absorbed into the normalization of $\tilde{f}_m$ --
see Eq.~(\ref{eqn:honorm}).

The condensate wavefunction can be obtained via the shooting and
Taylor expansion methods described in Appendix~\ref{app:numerics}.
Then Eqs.~(\ref{eqn:bog3})-(\ref{eqn:bog4}) can be solved
straightforwardly with standard numerical methods.  We use a
Laguerre discrete variable
representation~\cite{beck2000,mccurdy2003}, which is particularly
efficient for this geometry, allowing us to go to hundreds of basis
functions. The winding number $q$ of the Boguliubov modes was
checked for $q=0$ to $q=6$ over the entire domain of our study.

The results are shown in Fig.~\ref{fig:8}.  In panel (a), it is
apparent that a single ring soliton placed on top of the ground
state, i.e., with no central vortex, is always formally unstable to
the quadrupole, or $q=2$ mode. The instability time is given by $T=
-\omega/\mathrm{Im}(\Omega)$ in harmonic oscillator units. Typical
trap frequencies range from $\omega=2\pi\times 10$ Hz to $2\pi\times
100$ Hz.  Therefore, when $|\mathrm{Im}(\Omega)/\omega| \ll 1$, $T$
can be much longer than experimental time scales of 100 ms to 1 s.
In this case, we say that the solution is \emph{experimentally
stable}. The instability time $T$ can even be longer than the
lifetime of the condensate, the latter of which can range from 1 to
100 seconds. According to Fig.~\ref{fig:8}(a), this occurs for small
nonlinearities.  For larger nonlinearities, other modes, such as the
octopole ($q=3$), also become unstable.  In Fig.~\ref{fig:8}(b) is
shown the case of single ring soliton in the presence of a singly
quantized central vortex, i.e., $m=1$. The solution is again
formally unstable, though first to octopole rather than quadropole
perturbations.  For small nonlinearities it is experimentally
stable.

We did not quantitatively study solution stability for an infinite
well. However, we expect that the boundary provides additional
stability of ring solitons. To decay, ring solitons must break up
into pairs of vortices via a transverse oscillation.  To oscillate,
the ring soliton has to move away from the barrier and inwards
towards the origin.  This requires shrinking the circumference of
the ring, which costs energy, as a single ring feels an effective
potential which pushes it outwards, as for example in an unbounded
system.  We expect that very long decay times follow.  We make the
conjecture that, in the infinitely extended system, the presence of
an infinite number of rings, tightly pressed up against each other,
has the same effect with regards to the inner ring as the boundary
in the confined system.

\section{Discussion and Conclusions}
\label{sec:conclusions}

Soliton trains in one-dimensional BEC's, which are similar to the
nested ring solitons which form radial nodes, have been found to be
archetypes of planar soliton motions encountered in
three-dimensional BEC's. These solutions to the one-dimensional NLSE
are stationary states with islands of constant phase between equally
spaced nodes. When appropriately perturbed, they give rise to
soliton motion~\cite{reinhardt1997,carr2000a}. In fact, the
stationary solutions can be considered to be dark solitons in the
limiting case of zero soliton velocity, and the perturbation that
produces propagating ``gray'' solitons is the imposition of a slight
phase shift in the wavefunction across a node.  These
one-dimensional examples were found to have experimentally
accessible analogs in three-dimensional BEC's in which optically
induced phase shifts across a plane of symmetry resulted in soliton
motion~\cite{burger1999,denschlag2000,feder2000}. Correspondingly,
the two-dimensional ring solutions presented herein suggest the
possibility of creating ring soliton motion by imposing a phase
shift across the boundary of a disk.  We will address this in
subsequent work.

The existence of ring dark solitons has been predicted
theoretically~\cite{kivshar1994,dreischuh1996,frantzeskakis2000,nistazakis2001}
and demonstrated experimentally~\cite{neshev1997,dreischuh2002} in
the context of nonlinear optics.  It has been suggested that a
single ring dark soliton could be created in a confined
BEC~\cite{theocharis2003}.  A ring dark soliton corresponds to a
single node in our ring solutions.  It is known that a single ring
dark soliton in an infinitely extended system expands
indefinitely~\cite{kivshar1994}. This therefore clarifies why the
ring solutions require an infinite number of nodes in order to
obtain a stationary state.  It also explains why, for cylindrical
box boundary conditions, the creation of nodes tends to be towards
the edge of the condensate.  In Ref.~\cite{theocharis2003} it was
found that a single ring dark soliton was unstable to vortex pair
creation via the transverse, or snake instability in the near-field
($r\ll R_{\mathrm{TF}}$) Thomas-Fermi approximation in a harmonic
trap.  In agreement with this work, we have shown, without
approximations, that linear instability times can be made so long
that a ring soliton is in fact stable over the lifetime of the
experiment.  This result holds independent of the trap frequency in
the 2D plane.

The ring solutions that we have discussed in Sec.~\ref{sec:rings}
might be realized in an experiment that approximates a deep
cylindrical potential well, e.g. an optical trap using a
blue-detuned doughnut mode of a laser field~\cite{mclelland1991}. In
such a system, the ground state vortex solution will resemble that
of Fig.~\ref{fig:2}(a), where the abscissa is the radial coordinate
in units of the well radius. The first radially excited vortex
solution will then resemble that of Fig.~\ref{fig:2}(b). By use of
an optical phase-shifting technique such as that employed in
Refs.~\cite{burger1999,denschlag2000}, one might be able to generate
this state and observe its subsequent motion. We showed that the
same qualitative pattern of radial nodes that occurs for the
infinite well is also found in strongly oblate harmonic traps.

Concerning the central vortex core of the ring solutions, we note
that single vortices are quite long lived compared to experimental
time
scales~\cite{matthews1999,madison2000,aboshaeer2001,bretin2003}. It
is possible that forced excitation of the condensate may couple
resonantly, either directly or parametrically, to ring formation.
The same possibility exists for spherical shell solutions in the
observation of nodal spherical shells.  In two dimensions, unlike in
three, multiply charged vortices do not dynamically decay into
singly charged vortices with the addition of white noise to the
system~\cite{ivonin1999}, despite their being thermodynamically
unstable.  In fact, Pu {\it et al.} showed that stability regions
recur for large nonlinearity in two dimensions for
$m=2$~\cite{pu1999}.  Recent experiments have been able to create
and manipulate vortices of winding number greater than unity in a
variety of ways~\cite{aboshaeer2001,leanhardt2002b,engels2003}.
Therefore our study of vortices in two dimensions of winding number
higher than unity is experimentally relevant, despite their being
thermodynamically unstable~\cite{nozieres1990}.

In summary, we have elicited the form and properties of stationary
quantum vortices in Bose-Einstein condensates.  It was shown that
their axisymmetric stationary excitations take the form of nodal
rings, called ring solitons.  Quantization of these states can be
attained in confined geometries in two dimensions; we considered
both the infinite well and a harmonic trap.  Similar methods were
used to study the ground state and isotropic stationary excitations
in a spherical infinite well.  Two important aspects of these
solutions is that (a) the rings or spherical shells pile up near the
edge of the condensate, rather than being evenly spaced in $r$, in
contrast to the one-dimensional case, and (b) the chemical potential
depends linearly on the atomic interaction strength when the mean
field energy dominates over the kinetic energy, i.e., in the
Thomas-Fermi limit.  We showed that the ring solitons are
experimentally stable for weak nonlinearity.

This work was done in the same spirit as our previous articles on
the one-dimensional nonlinear Schr\"odinger
equation~\cite{carr2000a,carr2000b}.  A companion
work~\cite{carr2004k} treats the attractive case, which has features
radically different from the present study.  For instance, vortex
solutions are not monotonic in $r$. Moreover, there are a
denumerably infinite number of critical values of the determining
coefficient $a_0$ for fixed winding number which correspond to the
successive formation of nodes at $r=\infty$, even without an
external trapping potential.

Finally, we note that phenomena similar to the spherical shell
solutions have been experimentally observed in
BEC's~\cite{ginsberg2005}, while 2D BEC's appropriate to the
investigation of ring solitons are presently under intensive
investigation in experiments~\cite{stock2006,hadzibabic2006}.

We acknowledge several years of useful discussions with William
Reinhardt on the one-dimensional case, which provided the foundation
of our understanding of the two- and three-dimensional cases.  We
thank Joachim Brand for useful discussions.  LDC thanks the National
Science Foundation and the Department of Energy, Office of Basic
Energy Sciences via the Chemical Sciences, Geosciences and
Biosciences Division for support. The work of CWC was partially
supported by the Office of Naval Research and by the National
Science Foundation.

\appendix\section{Numerical Methods and Precision Issues} \label{app:numerics}

\subsection{Analytic Structure of the Solutions}
\label{ssec:power}

The following numerical methods are discussed explicitly for the
infinitely extended condensate, i.e., for a constant external
potential.  A slight modification for the infinite well is discussed
in App.~\ref{ssec:infinitewell}, and briefly for the strongly oblate
harmonic potential in Sec.~\ref{ssec:oblate}.

Since Eqs.~(\ref{eqn:gpe2d}) and~(\ref{eqn:gpe3d}) do not contain
any non-polynomial terms, one may begin with a power series solution
by a Taylor expansion around $\chi=0$ of the form \be
\eta_m(\chi)=\sum_{j=0}^{\infty} a_j \chi^{2j+m} \,
,\label{eqn:taylor} \ee where the $a_j$ are coefficients.  For
solutions which have the limiting behavior $\eta_m\rightarrow 0$ at
the origin, which is necessarily true for all non-divergent
solutions with $m\neq 0$, the nonlinear term $\eta_m^3$ becomes
negligible as $\chi\rightarrow 0$. Then the Bessel function
solutions to the linear Schr\"odinger equation are recovered.  This
will equally be true where $\eta$ has a node, in the neighborhood of
the node.  Thus near the origin the wavefunction must behave as
$\eta_m(\chi) \propto \chi^m$. This motivates the choice of the
exponent of $\chi$ in Eq.~(\ref{eqn:taylor}). By examination of
Eqs.~(\ref{eqn:gpe2d}) and~(\ref{eqn:gpe3d}), it is clear that only
even or odd powers of $\chi$ can have nonzero coefficients. The
Taylor series has been written in such a way as to eliminate all
terms which are obviously zero. Substituting Eq.~(\ref{eqn:taylor})
into Eqs.~(\ref{eqn:gpe2d}) and~(\ref{eqn:gpe3d}), the coefficients
can then be obtained recursively by equation of coefficients of
equal powers.  One finds that all coefficients $a_j$ for $j\neq 0$
can be expressed as a polynomial in odd powers of $a_0$ of order
$2(\lfloor j/(m+1)\rfloor)+1$, where $\lfloor x \rfloor$ denotes the
greatest integer less than or equal to $x$. For example, the first
few terms for $m=1$ are \bea
a_1&=&-\frac{1}{8}a_0\,,\nonumber\\
a_2&=&\frac{1}{192}(a_0+8a_0^3)\,,\nonumber\\
a_3&=&-\frac{1}{9216}(a_0+80a_0^3)\,,\nonumber\\
a_4&=&\frac{1}{737280}(a_0+656a_0^3+1152a_0^5)\,.\eea Thus the
coefficient $a_0$ is the only free parameter of the problem. We
consider only $a_0>0$, since for each solution $\eta_m(\chi;a_0)$,
there is a degenerate solution $\eta_m(\chi;-a_0)$.

The power series provides a useful, practical method for propagating
the solution of the NLSE away from the singular point at
$r=0$~\cite{ratio}.  However, it is not a practical method for
extension to large $r$ and we therefore use other methods in
intermediate and large $r$ regions. An asymptotic expansion which is
not formally convergent but nevertheless useful is obtained by the
transformation \be \zeta\equiv 1/\chi\, .\ee Then
Eq.~(\ref{eqn:gpe2d}) becomes \be
\left[\zeta^4\frac{\partial^2}{\partial\zeta^2}
+\zeta^3\frac{\partial}{\partial\zeta} -m^2\zeta^2
-\eta_m^2+1\right]\eta_m(\zeta)=0\, .\label{eqn:gpezeta}\ee  A
Taylor expansion around $\zeta=0$ yields the asymptotic power series
solution \bea \eta_1&=& 1-\frac{1}{2\,\chi^2}
-\frac{9}{8\,\chi^4}-\frac{161}{16\,\chi^6}
-\frac{24661}{128\,\chi^8}-\ldots\, , \\
\eta_2&=& 1-\frac{2}{\chi^2}
-\frac{6}{\chi^4}-\frac{68}{\chi^6}-\frac{1514}{\chi^8}-\ldots\, , \\
\eta_3&=& 1-\frac{9}{2\,\chi^2}
-\frac{153}{8\,\chi^4}-\frac{4473}{16\,\chi^6}
-\frac{962037}{128\,\chi^8}-\ldots\, , \eea {\it etc.} as
$\chi\rightarrow \infty$.  Since this series has no free parameters,
it is clear that only one value of the determining coefficient $a_0$
in Eq.~(\ref{eqn:taylor}) can lead to the vortex solution.  We
define this critical value as $a_0^{(m)}$. All values of
$a_0>a_0^{(m)}$ lead to divergent solutions, while values of
$a_0<a_0^{(m)}$ lead to solutions which asymptotically approach
zero, as shall be discussed in Sec.~\ref{sec:rings}.  It is in this
sense that the vortex solution manifests as a boundary between
divergent and non-divergent solutions.

Lastly, it is worthwhile to mention a limiting case which is useful
in Sec~\ref{sec:rings}.  For $m=0$ all coefficients $a_j$ are zero
except for $a_0$.  Examination of Eqs.~(\ref{eqn:gpe2d})
and~(\ref{eqn:gpezeta}) shows that the solution must be \be
\eta_0(\chi)=1\, .\ee This is the ground state in an extended system
in two dimensions.

\subsection{Solution by Pad\'e Approximant}
\label{ssec:pade}

It is desirable to determine the behavior of the vortex in
intermediate regions between zero and infinity.  The two point
Pad\'e approximant is defined by the rational function
\be
\frac{\mathcal{N}_{p,q}(\chi)}{\mathcal{D}_{p,q}(\chi)}=\frac{c_0 +
c_1 \chi+\cdots+c_{q-1}\chi^{q-1}}{d_0+d_1 \chi +\cdots+d_q \chi^q}
\, ,\ee where \be
\mathcal{N}_{p,q}(\chi)-f(\chi)\mathcal{D}_{p,q}(\chi)=\mathcal{O}(\chi^p)
\label{eqn:pade1}\ee as $\chi\rightarrow 0$ and \be
\mathcal{N}_{p,q}(\chi)-g(\chi)\mathcal{D}_{p,q}(\chi)=\mathcal{O}(\chi^{p-2q-1})
\label{eqn:pade2}\ee as $\chi\rightarrow \infty$ for all $p$ such
that $0\leq p \leq 2q$, with $p$ and $q$ integers.  The functions
$f(\chi),\,g(\chi)$ are power series expansions of the same function
as $\chi\rightarrow 0,\,\infty$.  The solution of
Eqs.~(\ref{eqn:pade1}) and~(\ref{eqn:pade2}) for the power series
expansions of Sec.~\ref{ssec:power} leads to a determination of the
critical value of the determining coefficient $a_0^{(m)}$ and
therefore a solution of the NLSE valid over all space.

For instance, taking $q=3$, one finds \be \eta_1 \simeq
\frac{\sqrt{2}\,\chi+2\chi^2}{1+\sqrt{2}+2\chi^2}\,.\ee  The
approximation can be successively improved.  Taking $q=4$ one finds
\be \eta_1\simeq \frac{16\sqrt{62}\,\chi + 248
\chi^2+30\sqrt{62}\,\chi^3}{124 +31\sqrt{62}\,\chi
+248\chi^2++30\sqrt{62}\,chi^3}\, ,\ee and so on.  By solving
Eqs.~(\ref{eqn:pade1}) and~(\ref{eqn:pade2}) at successively higher
order, one obtains a convergent value of $a_0^{(m)}$.

However, in practice this procedure is limited in precision due to
the appearance of spurious roots as well as by computation time. Due
to the nonlinear nature of Eqs.~(\ref{eqn:pade1})
and~(\ref{eqn:pade2}), there are multiple values of $a_0$ which
satisfy them.  These roots become sufficiently close to each other
so as to mislead root-finding algorithms.  Roots close to
$a_0^{(m)}$ tend to produce solutions which are asymptotically
correct but have spurious non-monotonic behavior, i.e., wiggles in
intermediate regions.  There are two options for root finding of
large systems of coupled polynomial equations. One may find all
roots and test them one by one. However, the computation time
becomes prohibitive for higher order polynomials and large numbers
of simultaneous equations. Or, one may use Newton's method or some
other local root-finding algorithm to find the root closest to the
correct one for the previous lowest order. The appearance of
spurious roots then becomes a limiting factor.

\begin{table}[t]
\centering \caption{\label{tab:1}Results of 2-point Pad\'e
approximant for the determining coefficient $a_0^{(m)}$ of quantum
vortex stationary states. $a_0^{(m)}$ is the first nonzero
coefficient in the power series solution of the 2D NLSE under the
assumption of a single central vortex of winding number $m$, and
determines
all subsequent coefficients.\\
}
\begin{tabular}{|c|c|c|}
\hline
Winding number $m$ & $a_0^{(m)}$ & Precision \\
\hline
0 &  1 & $\infty$\\
1 &  0.583189 & 6\\
2 &  0.15309 & 5\\
3 &  0.02618 & 4\\
4 &  0.00333 & 3\\
5 &  0.0002 & 1\\
\hline
\end{tabular}
\end{table}

\begin{table}
  \centering
  \caption{The best possible converged values via
  numerical shooting methods of the
  determining coefficient for quantum vortex stationary
  states.}\label{tab:2}
\begin{tabular}{|c|c|c|}
  \hline
  Winding number $m$ & $a_0^{(m)}$ & Precision \\
  \hline
  0 & 1 & $\infty$ \\
  1 & 0.5831 8949 5860 3292 7968 & 20 \\
  2 & 0.1530 9910 2859 54 & 14 \\
  3 & 0.0261 8342 07 & 9 \\
  4 & 0.0033 2717 34 & 8 \\
  5 & 0.0003 3659 39 & 7 \\
  \hline
\end{tabular}
\end{table}

In Table~\ref{tab:1} is shown the best convergent values of
$a_0^{(m)}$ for winding number zero to five.  Higher $m$ leads to
the appearance of more spurious roots and therefore a lower maximum
precision.  This may be understood as follows.  The coefficients in
the power series defined by Eq.~(\ref{eqn:taylor}) were polynomials
in $a_0$ of order $2(\lfloor j/(m+1)\rfloor)+1$.  High winding
number therefore requires a greatly increased number of terms in
order to obtain improved values of $a_0^{(m)}$. The higher the
number of terms, the greater the possibility that spurious roots
will appear. In practice, $q\simeq 30$ is the highest order
two-point Pad\'e approximant that is computable for vortex solutions
to the NLSE.

In the next section, we will demonstrate an alternative method that
does not suffer from the limitations of the two-point Pad\'e
approximant.  However, the Pad\'e approximant is worth retaining
because it provides interpolating functions in the form of rational
polynomials which can reproduce the small and large $\chi$ behavior
of the wavefunction to very high order.  We note that a thorough
treatment of the use of Pad\'e approximants in the study of vortex
and other solutions to the NLSE has been made by N.~G.
Berloff~\cite{berloff2004}.

\subsection{Solution by Numerical Shooting}
\label{ssec:shooting}

The 2D NLSE in the form given by Eq.~(\ref{eqn:gpe2d}) can be solved
by shooting. In this standard method~\cite{press1993}, one chooses
the values of $\eta(\chi_0)$ and $\eta'(\chi_0)$ for $\chi_0\ll 1$.
In this way, one can obtain an accurate relation between
$\eta(\chi_0)$ and $\eta'(\chi_0)$ via the power series of
Eq.~(\ref{eqn:taylor}). One then integrates the NLSE by initial
value methods towards arbitrarily large values of $\chi$. The
correct initial value of $\eta$ and $\eta'$ leads to the vortex
solution. Because the vortex solution lies on the boundary between
divergent and nondivergent behavior, it is quite easy to tell when
one has made a wrong choice: either $\eta$ diverges to infinity or
it oscillates and approaches zero. One chooses an initial value of
$a_0$, then iterates.  Note that the boundary cannot be chosen at
$\chi_0=0$, since $\eta=0$ is a valid solution to the 2D NLSE.
Rather, a value of $\chi_0$ which is exponentially small is used, so
as to ensure the accuracy of the power series solution.

In practice, the order of the power series is never a limiting
factor.  For example, we worked with 40 terms.  Up to ten additional
terms were tried without finding any difference in the results.
Instead, the two adjustable parameters in the calculation were
$\chi_0$ and the number of digits of internal precision used in our
numerical routine.  Of these, it was the latter that most strongly
affected the critical value of $a_0$.  In order to determine
$a_0^{(m)}$ to the highest possible precision, it was necessary to
use numbers of higher than double precision. It was found that 35
digits  was a practical maximum for our computing capabilities.  In
each case, the number of digits of precision was determined by
comparing the results using 32, 33, 34, and 35 digits of internal
computational precision.  In Table~\ref{tab:2} are shown the
results. Note that they are consistent with and greatly improve upon
those of the two-point Pad\'e approximant shown in
Table~\ref{tab:1}. As in Sec.~\ref{ssec:pade}, the calculations
proved more computationally difficult at higher winding number.
\begin{figure}
\centerline{\epsfxsize=8cm \leavevmode  \epsfbox{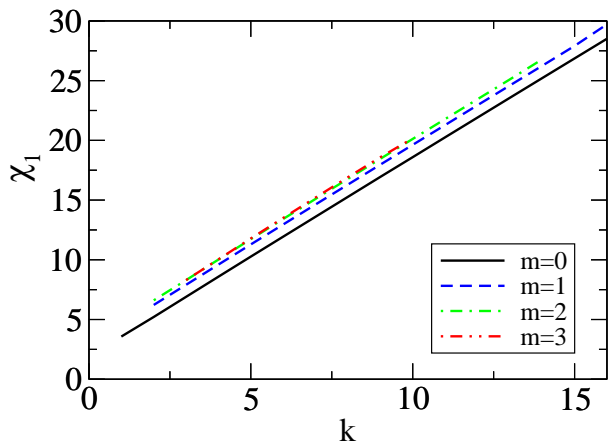}}
\caption{\label{fig:9} (color online) {\it Approaching the vortex
solution.} The position of the first node $\chi_1$ is shown as a
function of the number of digits of precision $k$ in the determining
coefficient $a_0^{(m)}$ for quantum vortex solutions to the 2D
nonlinear Schr\"odinger equation, as defined in
Eq.~(\ref{eqn:precision}). Observe that all curves are close to
parallel and {\it linear} in $k$, and as the winding number $m$
increases, the curves converge.  Note that all axes are
dimensionless.}
\end{figure}
The reader may ask why obtaining such high precision in the value of
$a_0^{(m)}$ is desirable. The reason is that each digit of precision
brings the solution a few units of $\chi$ closer to the exact vortex
solution.  In order to obtain a solution which is exactly on the
boundary between divergence and non-divergence an infinite number of
digits are required.  We illustrated this extremely sensitive
dependence of the determining coefficient on the number of digits of
precision in Fig.~\ref{fig:1}.  Figure~\ref{fig:1} depicts three
values of $a_0$ which approximate $a_0^{(m)}$ to $k$ digits of
precision, where $k$ is defined by \be
a_0=a_0^{(m)}(1-10^{-k})\,.\label{eqn:precision}\ee In
Eq.~(\ref{eqn:precision}), a small subtraction is made in the $k$th
digit, so that a convergent rather than divergent solution is
obtained.  In Fig.~\ref{fig:1}(a), the divergent solution is also
depicted for the same value of $k$, with $a_0=a_0^{(m)}(1+10^{-k})$:
i.e., a small \emph{addition} is made in the $k$th digit. It is in
this sense that the vortex solution is a boundary between convergent
and divergent solutions.

The figure shows the solution obtained via numerical shooting for
$k=16$, 8, and 2 in panels (a), (b), and (c), respectively.  A
winding number of $m=1$ is assumed.  In (b) and (c), the usual
Bessel function solution to the two-dimensional linear Schr\"odinger
equation is shown for comparison. One sees that the higher the
precision, the further the first node is pushed out towards large
values of $\chi$.  To move the node to infinity, an infinite number
of digits of precision is required.  All of the solutions except the
divergent one depicted in Fig.~\ref{fig:1}(a) are examples of ring
soliton solutions, as discussed in Sec.~\ref{sec:rings}.

One finds an intriguing relationship between the position of the
first node and the number of digits of precision $k$ in the critical
determining coefficient $a_0^{(m)}$, where $k$ is defined by
Eq.~(\ref{eqn:precision}). In Fig.~\ref{fig:9} is shown the position
of the first node $\chi_1(k)$. One observes that the relationship is
linear. For all values of the winding number but $m=0$, the curves
lie nearly on top of each other, and all are parallel. Clearly, from
Eq.~(\ref{eqn:gpe2d}), for large $\chi$ the term which depends on
$m$ becomes negligible. Note that the use of the special case $m=0$
ensures that, in at least one case, the exact value of $a_0^{(m)}$
is known. The best values of $a_0^{(m)}$ for $m\neq 0$ are given in
Table~\ref{tab:2}.

\subsection{Modified Numerical Method for the Infinite Well}
\label{ssec:infinitewell}

In order to quantize the solutions in an infinite well in two
dimensions, one holds the normalization and the cylinder radius to
be constant. The form of the wavefunction and the chemical potential
can be obtained as follows.  One calculates the dependence of the
normalization on $a_0$ as \be \int_0^{\chi_{j}(a_0)} d\chi\,\chi
[\eta_m(\chi)]^2 = \mathcal{N}_j\,, \label{eqn:normchi}\ee where
$\chi_{j}(a_0)$ is the distance to the $j$th node in $\eta(\chi)$.
As evident in Figs.~\ref{fig:1} and~\ref{fig:9}, a more useful
variable to determine the dependence of $\mathcal{N}_j$ on $a_0$ is
the number of digits of precision $k$, as defined by
Eq.~(\ref{eqn:precision}). Note that $k$ is not restricted to an
integer value.  The value of $k$ also determines $\chi_{j}$. From
Eq.~(\ref{eqn:efflength}), the chemical potential scaled to the
energy associated with the cylinder radius is \be \mu_{Rj}\equiv
\frac{2MR^2}{\hbar^2}\mu_j=\chi_j^2 \,\label{eqn:defmuj}\ee where
$\mu_j$ is the chemical potential for the $(j-1)$th excited state,
with $j=1$ giving the ground state. The function
$\mu_{Rj}(\mathcal{N}_j)$ can be calculated from
Eqs.~(\ref{eqn:normchi}) and~(\ref{eqn:defmuj}). This gives the
chemical potential as a function of the atomic interaction strength.

Quantized modes for the infinite spherical well in three dimensions
are calculated by similar methods.


\begin{thebibliography}{10}

\bibitem{pitaevskii1961}
L.~P. Pitaevskii, Sov. Phys. JETP {\bf 13},  451  (1961).

\bibitem{gross1961}
E.~P. Gross, Nuovo Cimento {\bf 20},  454  (1961).

\bibitem{fetter1972}
A.~L. Fetter, Ann. Phys. {\bf 70},  67  (1972).

\bibitem{donnelly1991}
R.~J. Donnelly, {\em Quantized Vortices in Helium II} (Cambridge
University
  Press, New York, 1991).

\bibitem{anderson1995}
M.~H. Anderson, {\it et al.}
  Science {\bf 269},  198  (1995).

\bibitem{davis1995}
K.~B. Davis {\it et al.}, Phys. Rev. Lett. {\bf 75},  3969  (1995).

\bibitem{bradleyCC1995}
C.~C. Bradley, C.~A. Sackett, J.~J. Tollett, and R.~G. Hulet, Phys.
Rev. Lett.
  {\bf 75},  1687  (1995).

\bibitem{bradleyCC1997}
C.~C. Bradley, C.~A. Sackett, and R.~G. Hulet, Phys. Rev. A {\bf
55},  3951
  (1997).

\bibitem{dalfovo1999}
F. Dalfovo, S. Giorgini, L.~P. Pitaevskii, and S. Stringari, Rev.
Mod. Phys.
  {\bf 71},  463  (1999).

\bibitem{leggett2001}
A.~J. Leggett, Rev. Mod. Phys. {\bf 73},  307  (2001).

\bibitem{greiner2003}
M. Greiner, C.~A. Regal, and D.~S. Jin, Nature {\bf 426},  437
(2003).

\bibitem{jochim2003b}
S. Jochim {\it et al.} Science {\bf 302},  2102  (2003).

\bibitem{zwierlein2003}
M.~W. Zwierlein {\it et al.} Phys. Rev. Lett. {\bf 91},  250401
(2003).

\bibitem{agrawal1995}
G.~P. Agrawal, {\em Nonlinear Fiber Optics}, 2nd ed. (Academic
Press, San
  Diego, 1995).

\bibitem{matthews1999}
M.~R. Matthews {\it et al.}, Phys. Rev. Lett. {\bf 83},  2498
(1999).

\bibitem{madison2000}
K.~W. Madison, F. Chevy, W. Wohlleben, and J. Dalibard, Phys. Rev.
Lett. {\bf
  84},  806  (2000).

\bibitem{burger1999}
S. Burger {\it et al.}, Phys. Rev. Lett. {\bf 83},  5198  (1999).

\bibitem{denschlag2000}
J. Denschlag {\it et al.}, Science {\bf 287},  97  (2000).

\bibitem{ruprecht1995}
P.~A. Ruprecht, M.~J. Holland, K. Burnett, and M. Edwards, Phys.
Rev. A {\bf
  51},  4704  (1995).

\bibitem{fetter2001}
A.~L. Fetter and A.~A. Svidzinsky, J. Phys.: Condens. Matter {\bf
13},  R135
  (2001).

\bibitem{feder1999}
D.~L. Feder, C.~W. Clark, and B.~I. Schneider, Phys. Rev. A {\bf
61},
  011601(R)  (1999).

\bibitem{williams1999}
J.~E. Williams and M.~J. Holland, Nature {\bf 401},  568  (1999).

\bibitem{theocharis2003}
G. Theocharis {\it et al.}, Phys. Rev. Lett. {\bf 90},  120403
(2003).

\bibitem{kivshar1994}
Y.~S. Kivshar and X. Yang, Phys. Rev. E {\bf 49},  1657  (1994).

\bibitem{dreischuh1996}
W. Dreischuh {\it et al.}, App. Phys. B
  {\bf 62},  139  (1996).

\bibitem{frantzeskakis2000}
D.~J. Frantzeskakis and B.~A. Malomed, Phys. Lett. A {\bf 264},  179
(2000).

\bibitem{nistazakis2001}
H.~E. Nistazakis, D.~J. Frantzeskakis, B.~A. Malomed, and P.~G.
Kevrekidis,
  Phys. Lett. A {\bf 285},  157  (2001).

\bibitem{neshev1997}
D. Neshev {\it et al.}, App. Phys. B {\bf 64},  429  (1997).

\bibitem{dreischuh2002}
A. Dreischuh {\it et al.}, Phys. Rev. E
  {\bf 66},  066611  (2002).

\bibitem{optics}
In the optics context, ring solitons are not confined by a harmonic
trap, which is the primary cause of their instability, as they can
expand without limit.

\bibitem{carr2000a}
L.~D. Carr, C.~W. Clark, and W.~P. Reinhardt, Phys. Rev. A {\bf 62},
063610
  (2000).

\bibitem{carr2000b}
L.~D. Carr, C.~W. Clark, and W.~P. Reinhardt, Phys. Rev. A {\bf 62},
063611
  (2000).

\bibitem{carr2004k}
L.~D. Carr and C.~W. Clark, Phys. Rev. Lett. {\bf 97}, 010403
(2006).

\bibitem{saffman1992}
P.~G. Saffman, {\em Vortex Dynamics} (Cambridge Univ. Press, New
York, 1992).

\bibitem{petrov2000}
D.~S. Petrov, M. Holzmann, and G.~V. Shlyapnikov, Phys. Rev. Lett.
{\bf 84},
  2551  (2000).

\bibitem{olshanii1998}
M. Olshanii, Phys. Rev. Lett. {\bf 81},  938  (1998).

\bibitem{petrov2000b}
D.~S. Petrov, G.~V. Shlyapnikov, and J.~T.~M. Walraven, Phys. Rev.
Lett. {\bf
  85},  3745  (2000).

\bibitem{carr2000e}
L.~D. Carr, M.~A. Leung, and W.~P. Reinhardt, J. Phys. B: At. Mol.
Opt. Phys.
  {\bf 33},  3983  (2000).

\bibitem{bongs2001}
K. Bongs {\it et al.}, Phys. Rev. A {\bf 63},  031602  (2001),
e-print cond-mat/0007381.

\bibitem{ketterle1999}
W. Ketterle, D.~S. Durfee, and D.~M. Stamper-Kurn,  in {\em
Bose-Einstein condensation in Atomic Gases}, edited by M. Inguscio,
S. Stringari, and C. E. Wieman, \emph{Proceedings of the
International School of Physics ``Enrico Fermi''} (IOS Press,
Amsterdam; Washington, D.C., 1999), pp.\ 67--176.

\bibitem{abramowitz1964}
{\em Handbook of Mathematical Functions}, edited by M. Abramowitz
and I.~A.
  Stegun (National Bureau of Standards, Washington, D. C., 1964).

\bibitem{dodd1996a}
R.~J. Dodd, J. Research NIST {\bf 101},  545  (1996).

\bibitem{pu1999}
H. Pu, C.~K. Law, J.~H. Eberly, and N.~P. Bigelow, Phys. Rev. A {\bf
59},  1533
   (1999).

\bibitem{brack1997}
M. Brack and R.~K. Bhaduri, {\em Semiclassical physics}
(Addison-Wesley, Reading, Massachusetts, 1997).

\bibitem{landau1977}
L.~D. Landau and E.~M. Lifshitz, {\em Quantum Mechanics
(Non-relativistic
  Theory)} (Pergamon Press, Tarrytown, New York, 1977), Vol.~3.

\bibitem{langer1937}
R.~E. Langer, Phys. Rev. {\bf 51},  669  (1937).

\bibitem{berry1972}
M.~V. Berry and K.~E. Mount, Rep. Prog. Phys. {\bf 35},  315
(1972).

\bibitem{svidzinsky1998}
A.~A. Svidzinsky and A.~L. Fetter, Phys. Rev. A {\bf 58},  3168
(1998).

\bibitem{garay2001}
L.~J. Garay, J.~R. Anglin, J.~I. Cirac, and P. Zoller, Phys. Rev. A
{\bf 63},
  023611  (2001).

\bibitem{beck2000}
M.~H. Beck, A. J\"ackle, G.~A. Worth, and H.-D. Meyer, Phys. Rep.
{\bf 324},  1
   (2000).

\bibitem{mccurdy2003}
C.~W. McCurdy, W.~A. Isaacs, H.-D. Meyer, and T.~N. Rescigno, Phys.
Rev. A {\bf
  67},  042708  (2003).

\bibitem{reinhardt1997}
W.~P. Reinhardt and C.~W. Clark, J. Phys. B: At. Mol. Opt. Phys.
{\bf 30},
  L785  (1997).

\bibitem{feder2000}
D.~L. Feder {\it et al.},
  Phys. Rev. A {\bf 62},  053606  (2000).

\bibitem{mclelland1991}
J.~J. McClelland and M.~R. Scheinfein, J. Opt. Soc. Am. B {\bf 8},
1974
  (1991).

\bibitem{aboshaeer2001}
J.~R. Abo-Shaeer, C. Raman, J.~M. Vogels, and W. Ketterle, Science
{\bf 292},
  476  (2001).

\bibitem{bretin2003}
V. Bretin {\it et al.}, Phys.
  Rev. Lett. {\bf 90},  100403  (2003).

\bibitem{ivonin1999}
I.~A. Ivonin, V.~P. Pavlenko, and H. Persson, Phys. Rev. E {\bf 60},
492
  (1999).

\bibitem{leanhardt2002b}
A.~E. Leanhardt {\it et al.}, Phys. Rev. Lett. {\bf 89},  190403
(2002).

\bibitem{engels2003}
P. Engels {\it et al.},
  Phys. Rev. Lett. {\bf 90},  170405  (2003).

\bibitem{nozieres1990}
P. Nozi\`eres and D. Pines, {\em The Theory of Quantum Liquids}
  (Addison-Wesley, New York, 1990), Vol.~II.

\bibitem{ginsberg2005}
N.~J. Ginsberg, J. Brand, and L.~V. Hau, Phys. Rev. Lett. {\bf 94},
040403
  (2005).

\bibitem{stock2006}
S. Stock {\it et al.}, Physical
  Review Letters {\bf 95},  190403  (2005).

\bibitem{hadzibabic2006}
Z. Hadzibabic {\it et al.}, Nature in press, e-print
cond-mat/0605291 (2006).

\bibitem{ratio}
Numerical evaluation of the ratios $a_{n+1}/a_n$ up to $n=40$
indicates that
  the radius of convergence approaches infinity as $a_0\rightarrow 0$, while
  for $a_0>0$ the radius of convergence appears to be finite.

\bibitem{berloff2004}
N.~G. Berloff, J. Phys. A: Math. Gen. {\bf 37},  1617  (2004).

\bibitem{press1993}
W.~H. Press, S.~A. Teukolsky, W.~T. Vetterling, and B.~P. Flannery,
{\em
  Numerical Recipes in C: The Art of Scientific Computing} (Cambridge Univ.
  Press, Cambridge, U.K., 1993).

\end{thebibliography}

\end{document}